**Fears about AI-mediated communication are grounded in different expectations for one's own versus others' use**

Zoe A. Purcell[1*], Mengchen Dong[2], Anne-Marie Nussberger[2], Nils Köbis[2], & Maurice Jakesch[3]

1. Artificial and Natural Intelligence Institute of Toulouse (ANITI), University of Toulouse, Toulouse, France

2. Center for Humans and Machines, Max Planck Institute for Human Development, Berlin, Germany

3. Cornell Tech, Cornell University, New York, USA

*Corresponding author: zoe.purcell@iast.fr

**Abstract**

The rapid development of AI-mediated communication technologies (AICTs) – digital tools that use AI to augment our interpersonal messages – has created concerns about the future of interpersonal trust and prompted urgent discussions about disclosure and uptake. We contribute to this discussion by assessing perceptions about acceptability and use of open and secret AICTs for oneself and others. In two studies with representative samples (UK: $N$=477, US: $N$=765), we found (a) that secret (i.e., undisclosed) AICT use is deemed less acceptable than open AICT use, (b) people overestimate others' AICT use, and (c) people expect others to use AICTs irresponsibly. Thus, we raise concerns about the potential for misperceptions and different expectations for others to drive self-fulfilling pessimistic outlooks about AI-mediated communication.

**Layperson Abstract**

AI-mediated communication technologies (AICTs) can augment our messages in real-time – making people's text more persuasive, voices sound happier, and images look more attractive. However, as these technologies improve (e.g., with large language models like GPT-4) distinguishing between original and augmented communications is getting harder and harder. This emergence of powerful AICTs has sparked concerns about their potential to enable widespread deception and cause large-scale degradation of trust in interpersonal communication. We assess the validity of those concerns by examining the acceptability of open/secret AICT use and people's expectations about their own versus others' use. We confirm the pessimistic hypotheses; secret AICTs were rated as less acceptable than open AICTs, people overestimated others' AICT use, and people expected other's AICT use to be irresponsible. We discuss how misperceptions and different expectations for others' AICT use may drive self-fulfilling, dystopian outlooks when individuals and regulators act according to them.

## Keywords

**Fears about AI-mediated communication are grounded in different expectations for one's own versus others' use**

Humans are social beings – our lives and identities are defined by our relationships. Our relationships depend on communication, assumptions of authenticity, and interpersonal trust (Grueter & White, 2014; Hruschka, 2010). Artificial Intelligence (AI), especially in the form of AI-mediated communication technologies (AICTs), is radically transforming the way we communicate (Hancock et al., 2020; Sundar, 2020; Sundar & Lee, 2022). AICTs have the potential to improve the efficacy of our communication. However, they also have the potential to increase deception, threaten our perceptions of others' authenticity, and promote mistrust (Jago, 2019; Jakesch, Hancock, et al., 2023). These factors may form a dystopian vision for the future of AICTs and human-AI interaction. However, they rest on assumptions that people believe that secret AICT use is problematic, that other people will use them, and that others' use will be irresponsible. We examine these assumptions hoping to find a way out of this pessimistic vision.

AI is now involved in many communication tools, such as chatbots like OpenAI's ChatGPT and voice assistants like Amazon's Alexa. While these AI-as-*agent* communication tools are important contributors to the social arena, our concerns lie with the more complicated addition of AI-as-*mediator* communication tools. AI-mediated communication is communication between *people* in which a computational agent modifies, augments or generates messages to accomplish communication or interpersonal goals (Hancock et al., 2020). AICTs include narrow-AI text editors like Grammarly and Wordtune, but also incorporate tools that lean on more general AI like large language models in GPT-4. AICTs produce communications where the line between what is generated by the human sender and what is generated by the AI contributor becomes increasingly blurred. As discussions about the ethics and implications of this complex technology emerge, it is critical that we understand the perceptions and individual outlooks that will guide those discussions and subsequent regulation.

AICTs are inherently divisive because of their potential to increase deception and mistrust (Jakesch et al., 2019; Köbis et al., 2021). Hence, they have sparked lively debates on the ethics and implications of emerging AI language technologies (Chesney & Citron, 2019; Hancock et al., 2020; Jakesch, Bhat, et al., 2023; Ruggeri, 2023). Regulators are scrambling to define, discourage and restrict the inappropriate use of AI-generation tools. For instance, the New York City school district was among the first to ban ChatGPT (Yang, 2023), Italy has imposed a blanket ban (Browne, 2023), and Europol has raised grave concerns about its criminal potential (Chee, 2023). Meanwhile, developers are seeking technological solutions, such as ChatGPT-detection algorithms like ZeroGPT. But the growing pervasiveness of AICTs will increasingly blur lines of ownership, authenticity, and agency, raising fundamental questions about AICTs' perception, uptake, and consequence concerning policymakers as much as scientific experts and common users (Hsu, 2023; Munch et al., 2023). While much funding goes into developing these tools, much less attention is placed on understanding their impact on individual outlooks, interpersonal trust, and societal ramifications.

In this article, we explore the factors that will guide individual outlooks towards AICTs and its uptake as they are invariably developed. We study critical perceptions around acceptability and expectations of AICT use, and pragmatic antecedents such as transparency – whether AICTs are used openly or secretly – and the user – whether we are considering our own- or others' AICT use. We focus on the self-other distinction because people's perceptions of (1) what behaviors are acceptable or not (i.e., injunctive norms) and (2) what most other people typically do (i.e., descriptive norms) are both crucial factors that predict a person's behaviors, especially those that are morally dubious (Bicchieri & Xiao, 2009).

Moreover, people do not impose identical moral standards on themselves versus others (Valdesolo & DeSteno, 2007; Weiss et al., 2018), nor in public versus private settings (Dana et al., 2007).

In this regard, we address three questions. Are secret AICTs perceived as less acceptable than open AICTs (Claim 1)? Do people expect others to use AICTs more than they would themselves (Claim 2)? And, do people expect others' use to be irresponsible (Claim 3)? Beyond these questions, we explored the roles of individual factors including beliefs about- and familiarity with AICTs. Our findings have implications at the individual, interpersonal, and societal levels. The results shed light on whether individuals hold pessimistic outlooks about the future of AICTs and whether AICTs will affect interpersonal trust. At the societal level, our findings provide insights about the public position on AICTs and how this position will impact the development of technologies and policies concerning the broader human-AI relationship.

## Study 1

### *Method*

**Participants and method.** Study 1 examined Claim 1, whether acceptability would be impacted by the *transparency* of the tool use (secret or open) and Claim 2, whether usage expectations would be impacted by the *user* in question (self or other). A representative sample of UK participants ($N = 477$) aged 18 to 75 ($M = 48.48$, $SD = 15.83$; females = 266[1]) was presented with secret and open AICTs, and indicated expectations about their own versus others' AICT use. All participants saw six examples of AICTs: three mediums (video, voice, text) at two augmentation strengths (weak- and strong-augmentation; see Figure 1 and Supplementary Material for examples). For each AICT, every participant was asked four questions, two about *acceptability* 1) How acceptable do you think it is for someone to **openly** use this technology? 2) How acceptable do you think it is for someone to **secretly** use this technology? [1= Very unacceptable to 5 = Very acceptable], and two about *usage expectations* 1) How likely are **others** to use such technology? and 2) How likely are **you** to use such technology? [1 = Very unlikely to 5 = Very likely].

---

[1] Other genders: Male = 200, non-binary = 4, prefer not to say = 3, no response = 4.

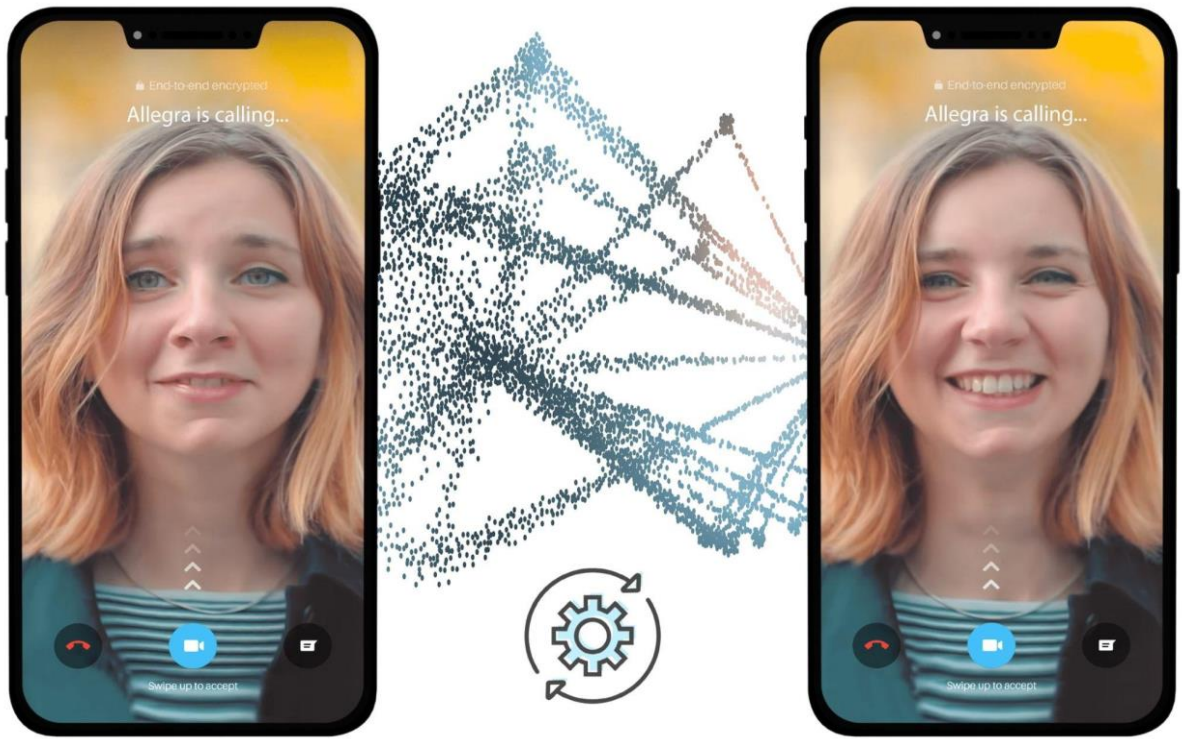

Figure 1. *An example of the stimuli used in Study 1 (medium=video, strength=strong). See OSF for more examples.*

### Results

**Acceptability.** In line with Claim 1, we found that people are less accepting of secret than open AICT use (Figure 2 (A)). We used a linear mixed model predicting 'acceptability' from the AICT's transparency and included participant, medium, and strength were included as random intercept effects. The model indicated a significant main effect of transparency ($B$ = -0.69, 95% *CI* [-0.74, -0.64], $t(5718)$ = -26.53, $p < .001$). Thus, people were less accepting of secret use of AICTs (e$M$ = 2.72, $SE$ = .25) than open use (e$M$ = 3.41, $SE$ = .25).

**Usage expectations.** In line with Claim 2, we observed that people believed others would use AICTs more than they would themselves (Figure 2 (B)). We employed a linear mixed model predicting expected usage from user (self or other) and including participant, medium, and strength as random intercepts. The model showed a significant main effect of user ($B = 0.93$, 95% $CI$ [0.88, 0.98], $t(5718) = 33.98$, $p < .001$). Thus, people expected that others ($eM =$ 3.73, $SE = .19$) would use AICTs more than they would themselves ($eM = 2.80$, $SE = .19$).

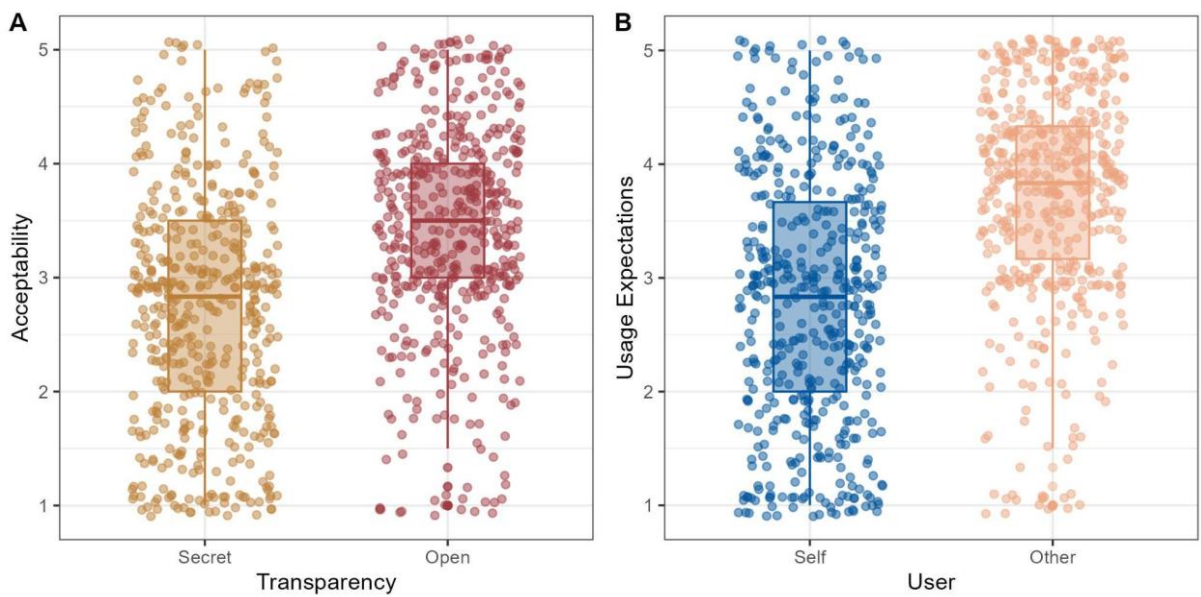

Figure 2. *The results from Study 1 show that participants were more accepting of open than secret use of AICTs (Panel A) and participants expected others to use AICTs more than they would themselves (Panel B). Data points are jittered 0.1 vertically.*

Study 1 found that, in support of Claim 1, participants were more accepting of open than secret AICT use, and that, in support of Claim 2, participants believed others were more likely to use AICTs than they were themselves. In Study 1, we observed how participants responded when transparency/user was salient. Namely, for each medium (text, voice or video), every participant was asked how acceptable it would be to use it *openly* and *secretly*, how much they expected to use these tools *themselves* and how much they expected *others* to use these tools. Hence, each participant was able to directly contrast their responses in regard to transparency and user. In addition to examining Claim 3, Study 2 examined the boundaries of- and the relationship between Claims 1 and 2; including whether these response patterns hold when transparency and user are not made salient.

**Study 2**

All methods and analyses for Study 2 were preregistered at OSF.

*Method*

**Participants and method.** Study 2 assessed whether people expect others to use AICTs responsibly by examining the relationship between perceived acceptability and usage expectations for own vs. others' AICT use (Claim 3). Additionally, by manipulating transparency and user between-subjects, Study 2 allowed us to examine the boundary conditions for Claims 1 and 2.

We surveyed a representative sample of the US population, stratified across age, sex and ethnicity ($N = 765$) aged 18 to 93 ($M = 45.47$, $SD = 15.75$; females = 375[2]). We implemented a 2 (transparency: open vs. secret; between-subjects) by 2 (user: self vs. other; between-subjects) by 3 (medium: text vs. voice vs. video; within-subjects) design. Participants were allocated to one of four conditions such that they saw examples of AICTs that were either open *or* secret, thus manipulating transparency. These were for use by the participant themselves *or* by others, thus manipulating user. Each participant saw three examples of AICTs: text, voice, and video. To contain the number of comparisons, only 'strong' augmentation AICTs were used (see Figures 1 and 3 and Supplementary Information for examples).

---

[2] Other genders: Male = 379, non-binary = 8, prefer not to say = 3.

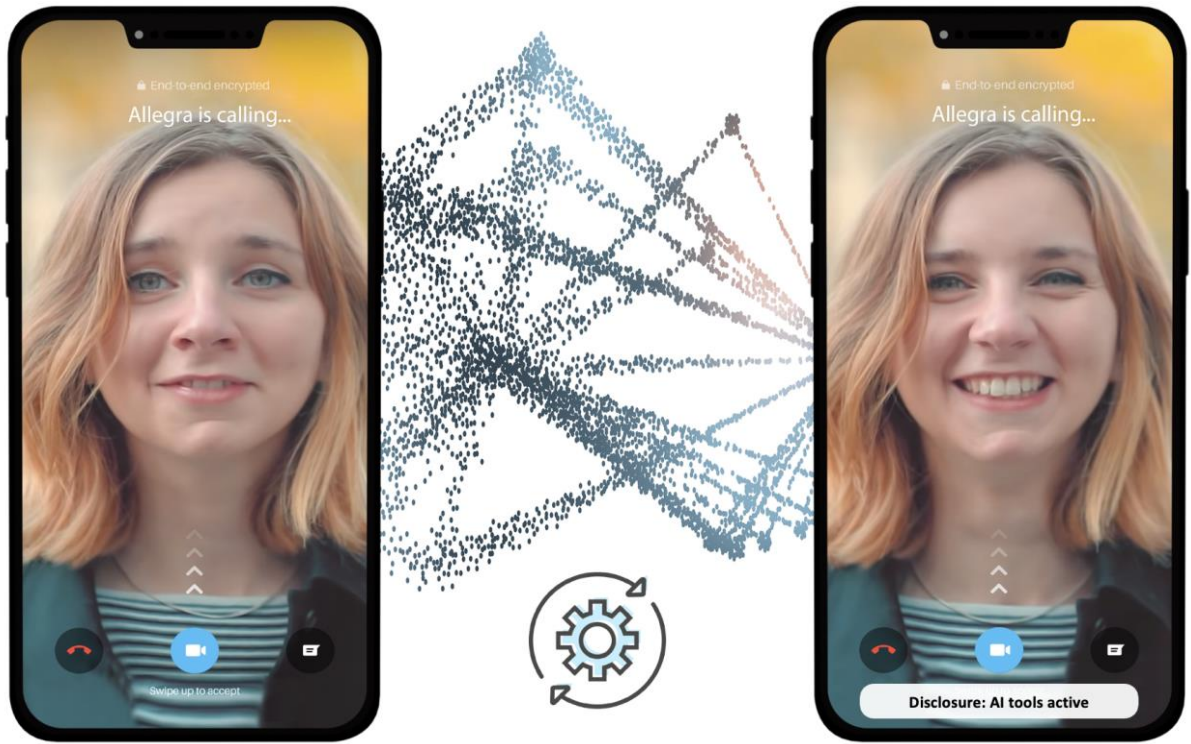

Figure 3. *An example of the stimuli used in Study 2 (transparency=open, medium=video). For an example of the equivalent stimuli in the transparency=secret condition, see Figure 1.*

For each of the three AICTs, participants were asked about usage expectations [0 = Very Unlikely to 100 = Very Likely] and acceptability [0 = Very Unacceptable to 100 = Very Acceptable]. For example, a participant in the 'open-self' condition was presented with three AICTs that a) declare (visually or audibly) that AI assistance tools have been/are being used and b) was asked about whether they would use the AICT and c) how acceptable it is to do so. Whereas a participant in the 'secret-other' condition was presented with three AICTs that a) do not declare whether AI assistance tools are involved b) would be asked about whether others would use the AICT and c) how acceptable it would be to do so.

For each medium, we explored several potentially related individual factors: 1) the respondent's familiarity with AICTs (e.g., "How familiar are you with AI-powered video editing tools that alter one's appearance during a video call? [0=Not familiar at all to 100 = Extremely familiar]"),  2) whether participants believed AICT use would lead to a loss of information (e.g., "How much information is lost if a person uses AI-powered video editing tools to alter their appearance during a video call? [0= None at all to 100 = A great deal]"), and 3) whether they perceived AICT use or non-use as indicative of another person's character (e.g., "Whether or not another person uses AI-powered video editing tools tells me something about that other person's character. [0=Strongly disagree to 100 = Strongly agree]"). We report results of individual factors and how they relate to the primary claims in the Appendix.

### Results

The following section addresses 1) ratings of acceptability and the boundaries of Claim 1, 2) ratings of usage expectations and the boundaries of Claim 2 and finally, 3) Claim 3 – expectations about irresponsible use and the relationship between ratings of acceptability and usage expectations.

**Acceptability.** Regarding Claim 1, under these non-salient conditions, transparency had a small and marginally significant effect on acceptability. However, whether the participant was assessing the acceptability of their *own* use or that of *others* affected ratings of AICT acceptability, people rated AICTs as more acceptable when considering others' use than when considering their own use.

We used a linear mixed model predicting 'acceptability' from user (self, other), transparency (open, secret), and their interaction with participant and medium included as random intercept effects. In the model predicting acceptability, the main effect of user was significant ($B = 4.79$, 95% $CI$ [1.31, 8.26], $t(2288) = 2.70$, $p = .007$), people were less accepting when considering their own use ($eM = 52.85$, $SE = 4.94$) than that of others' ($eM = 57.64$, $SE = 4.94$). Regarding Claim 1, the main effect of transparency was marginal ($B = 3.25$, 95% $CI$ [-0.23, 6.73], $t(2288) = 1.83$, $p = .067$), there was only a slight difference in their acceptance of open ($eM$ = 56.87, $SE$ = 4.94) compared to secret AICTs ($eM$ = 53.62, $SE$ = 4.94). The interaction of transparency and user on acceptability was not significant ($B$ = -1.74, 95% $CI$ [-8.69, 5.22], $t$(2288) = -0.49, $p$ = .624).

**Usage expectations.** In line with Claim 2, people believed others would use AICTs more than they would themselves, even in these non-salient conditions. However, transparency did not affect usage expectations.

We used a linear mixed model predicting 'usage' from user (self, other), transparency (open, secret), and the user-transparency interaction with participant and medium included as random intercept effects. In line with Claim 2, the main effect of user was significant ($B$ = 30.97, 95% $CI$ [27.58, 34.36], $t$(2288) = 17.92, $p$ < .001). People expected that others ($eM$ = 71.41, $SE$ = 4.10) would use AICTs more than they would themselves ($eM$ = 40.44, $SE$ = 4.10). However, the main effect of transparency was not significant as a main effect ($B$ = 0.47, 95% $CI$ [-2.92, 3.86], $t$(2288) = 0.27, $p$ = .787), people did not have different usage expectations for open ($eM$ = 56.16, $SE$ = 4.10) compared to secret AICTs ($eM$ = 55.70, $SE$ = 4.10). The interaction between transparency and user was also not significant ($B$ = -0.85, 95% $CI$ [-7.63, 5.92], $t$(2288) = -0.25, $p$ = .805).

**Usage expectations and acceptability.** As illustrated in Figure 4, we found support for Claim 3: the relationship between acceptability and usage was stronger for participants considering their own use than that of others.

To analyze this finding, we used a linear mixed effects model predicting rating scores from user, transparency, and evaluation (acceptability, usage expectations), as well as their two- and three-way interactions. Participant and medium were included as random intercept effects. In line with Claim 3, we found a significant interaction between user and evaluation ($B$ = 26.18, 95% $CI$ [23.62, 28.75], $t$(4579) = 20.00, $p$ < .001). Interestingly, we also observed a smaller but significant interaction between transparency and evaluation ($B$ = -2.78, 95% $CI$ [-5.35, -0.22], $t$(4579) = -2.13, $p$ = 0.034). These effects were not qualified by a three-way interaction ($p$ = .736; see Appendix for other effects in the model).

We explored the user-by-evaluation interaction by examining the relationship between acceptability and usage expectations separately for participants considering their *own* AICT use (user=self) and for participants considering *others'* AICT use (user=other). We used a linear mixed model predicting usage expectations from acceptability scores with participant and medium as random effects. The relationship between acceptability and usage expectations was much stronger for 'self' AICT use ($B$ = 0.78, 95% $CI$ [0.73, 0.82], $t$(1126) = 35.82, $p$ < .001) than for 'other' AICT use ($B$ = 0.34, 95% $CI$ [0.30, 0.38], t(1159) = 15.56, $p$ < .001).

Similarly, to explore the transparency-by-evaluation interaction, we examined the relationship between acceptability and usage expectations separately for participants considering open AICTs and those considering secret AICTs. We used a linear mixed model predicting usage expectations from acceptability scores with participant and medium as random effects. However, the relationship between acceptability and usage expectations was similar for open AICTs ($B$ = 0.59, 95% $CI$ [0.54, 0.63], $t$(1129) = 23.81, $p$ < .001) and secret AICTs ($B$ = 0.58, 95% $CI$ [0.53, 0.63], $t$(1156) = 23.26, $p$ < .001).

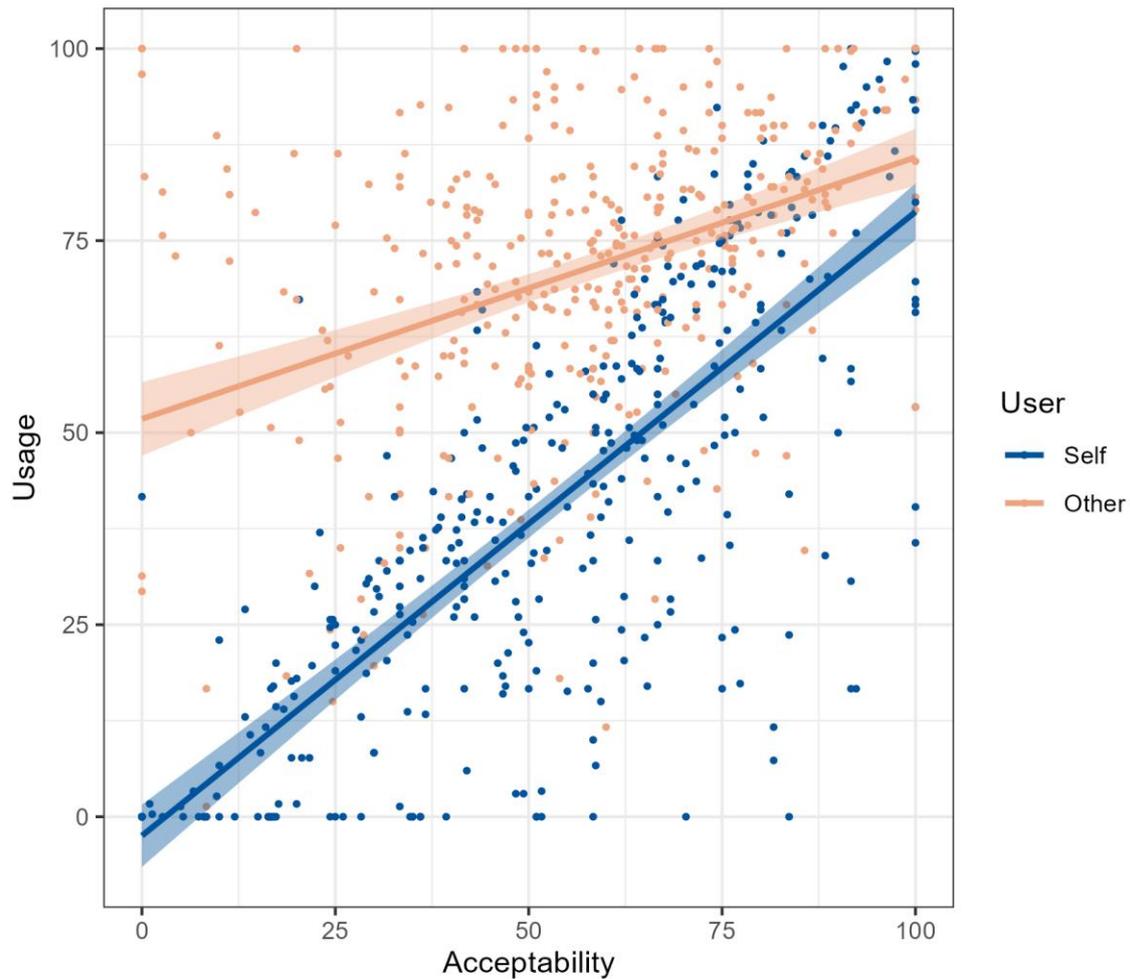

Figure 4. *The results from Study 2 show that the relationship between acceptability and usage expectations is stronger when people consider their own use than when they consider others' use.*

Study 2 found that, regarding Claim 1, under non-salient conditions, the effect of transparency on acceptability was only marginally significant; those participants rating open AICTs were only slightly more accepting than those rating secret AICTs. Regarding Claim 2, participants believed others were more likely to use AICTs than they were themselves – even in non-salient conditions. Study 2 also found that others' AICT use was deemed more acceptable than one's own AICT use, but transparency did not affect usage expectations.

Study 2 found clear support for Claim 3, the relationship between judgments of what is acceptable and expectations for use are less aligned when people consider others' use than when they consider their own. While people think that their own use will be responsible, in that it is in line with what they deem acceptable, they do not expect others' use to be as responsible.

## Discussion

We are witnessing exponential growth in the development and uptake of AI communication tools. One exemplar, Chat-GPT, gained one million users in just five days (Buchholz, 2023). Despite this proliferation, little is known about individual outlooks on this development nor, consequently, about how we can expect this development to evolve. In this regard, we studied people's acceptability of and expectations for AICT use and explored two key contextual factors: AICT transparency (open vs. secret use) and the user in question (self vs. others). Several untested possibilities might have challenged the assumption that people are concerned about the progress of AI-mediated communication: People may not have been concerned with secret AICT use; people may not have expected others to use AICTs; and if they did, then they might have believed that others would do so in a responsible way. For all three claims, our results confirmed the pessimistic hypothesis.

People rated secret AICTs – those that do not notify the receiver of AI involvement – as less acceptable than open AICTs – those that did notify the receiver of the AI involvement. This effect was powerful in Study 1 when the questions about open and secret use were presented simultaneously. Interestingly, this effect was weaker and only marginally significant in Study 2 where transparency was less salient. In Study 2, we deployed a between-subjects manipulation of open/secret use that did not explicitly point participants to these operation-modes for AICTs. These findings are important for understanding that people, in principle, care about transparency in AICT use. However, the acceptability of AI-use will depend on how the technologies are introduced and how transparency measures are being communicated. This insight is critical for future AI acceptability research. Moreover, regarding AICT-uptake and consumer preferences, these findings imply that transparency should be seriously considered in the ethical design of AICT products. But, that contextual factors like awareness of different design options will be critical to user decisions who, as their awareness increases, may gradually discard AICT products that are not tailored to their ethical preferences.

When considering own vs. others' AICT use, people showed robust self-other differences in acceptability judgments and usage expectations of AICTs. People expect others to use AICTs more than they expect to use them themselves. This finding suggests that people overestimate the extent to which others would use AICTs. Interestingly, this was accompanied by a belief that others' AICT use is more acceptable than one's own. Such diverging standards of acceptability could suggest that, for AICT use, others are held to different (lower) standards of acceptable behaviour or that, in line with the suggestion that behaviour drives justification, the expectation of more frequent AICT use by others could be driving up the perceived acceptability of that use (Eriksson et al., 2015). Similarly, people may anticipate that others have a greater need for or greater access to AICT tools than they do themselves.

A self-other distinction also emerged regarding the association between acceptability and usage. While people's expectations for their *own* use were strongly aligned with their perceptions of what was acceptable, this was less-so for their expectations for *others'* usage. This result suggests that while we expect our own AICT use to be responsible in so far as it reflects what we deem as acceptable, we do not expect others' AICT use to follow the same principles. Coupled with our finding that people overestimate others' AICT use, this insight suggests that people may hold broadly negative outlooks for the evolution of AICTs and AICT use such that they believe uptake will be greater for others, and not align with their individual beliefs about what is acceptable.

Our exploration of individual factors gives some indication of the possible drivers of acceptability and usage expectations. For example, regarding AICTs, what people deem as more acceptable is related to their beliefs about whether AICT use is indicative of the user's character and whether they believe AICT use yields a loss of information. Additionally, these findings may inform our interpretation of the self-other distinctions that emerge across our primary claims. It appears that people's perceptions of whether AICT use is 'signaling' – that is, whether it tells you something about the user's character – are strongly related to ratings of one's own but not others' AICT use. Indeed, when people believe AICT use can be morally negative, they show harsher standards on themselves than others to manage their reputation (Dong et al., 2023); hence, the lower reported expectations for own use could reflect a strategic move for reputation management. Hence, people's actual choice of AICT products may be guided by their (inaccurate) beliefs about others' usage – a similar pattern has been observed for other morally charged behavior (Dorrough et al., 2023).

AICT development and uptake are rapidly progressing and, with it, explosive debates about acceptability and regulation. Future AICT development, policy, and uptake will likely be guided by what the public deems acceptable and how it expects these technologies to be used. We provide empirical evidence for claims that acceptability and usage expectations are associated with transparency (open/secret) and the user (self/other). We offer insights into the role of a fundamental psychological distinction between perceptions of self and other and the influences of individual beliefs about AICTs. We find that – underpinned by the robust self-other distinction – people indeed hold negative outlooks regarding AICT development due to the overestimation of others' use and the belief that that use will be irresponsible. These findings validate claims that people are fearful about AICTs and suggest that this fear is grounded in different expectations for one's own and others' use. In the future, misperceptions subsequent fears may be alleviated by increased awareness/familiarity or open-use AICT policies (especially given people's preference for open over secret AICTs) but, left unchallenged, they may be self-fulfilling by expediting negative individual outlooks, degrading interpersonal trust, and limiting the sound development of technologies and regulations that impact the human-AI interaction.

**Appendix**

**Extended Results for Study 2**

**Acceptability and usage expectations.** To examine the relationship between ratings of acceptability and usage expectations, and their interactions with transparency and user, we used a linear mixed model predicting rating scores from user, transparency, and 'evaluation' (acceptability or usage expectations). Participant and medium were included as random effects. The model's intercept, corresponding to the model mean, is at 55.59 (95% CI [47.35, 63.83], $t(4579) = 13.23$, $p < .001$). Within this model:

- The effect of transparency was statistically non-significant and positive ($B = 1.86$, 95% CI [-1.24, 4.96], $t(4579) = 1.18$, $p = 0.239$).

- The effect of user was statistically significant and positive ($B = 17.88$, 95% CI [14.78, 20.98], $t(4579) = 11.31$, $p < .001$;

- The effect of evaluation was statistically non-significant and positive ($B = 0.68$, 95% CI [-0.60, 1.97], $t(4579) = 1.04$, $p = 0.297$).

- The effect of the transparency × user interaction was statistically non-significant and negative ($B = -1.30$, 95% CI [-7.49, 4.90], $t(4579) = -0.41$, $p = 0.682$; see below for simple effects.

- The interaction effect of transparency × evaluation was statistically significant and negative ($B = -2.78$, 95% CI [-5.35, -0.22], $t(4579) = -2.13$, $p = 0.034$); see main text for more details.

- The effect of the user × evaluation interaction was statistically significant and positive ($B = 26.18$, 95% CI [23.62, 28.75], $t(4579) = 20.00$, $p < .001$); see main text for more details.

- The three way interaction effect of transparency × user × evaluation was statistically non-significant and positive ($B$ = 0.88, 95% CI [-4.25, 6.02], $t$(4579) = 0.34, $p$ = 0.736).

**Individual factors.**

In addition to our primary analyses, we examined how three factors – familiarity, information loss, and character signal – impacted the acceptability of AICTs and people's usage expectations.

**Individual factors and acceptability.** We found that greater acceptability was associated with a) greater familiarity with AICTs, b) lower expectations of information loss due to AICT use, and c) weaker character signal provided by AICT use (see Figure A). These results were confirmed using three linear mixed effects models with the same structure as for usage expectations (see below). Additionally, we found that the effect of information loss on acceptability was stronger for secret than open AICTs. No other effects reached statistical significance.

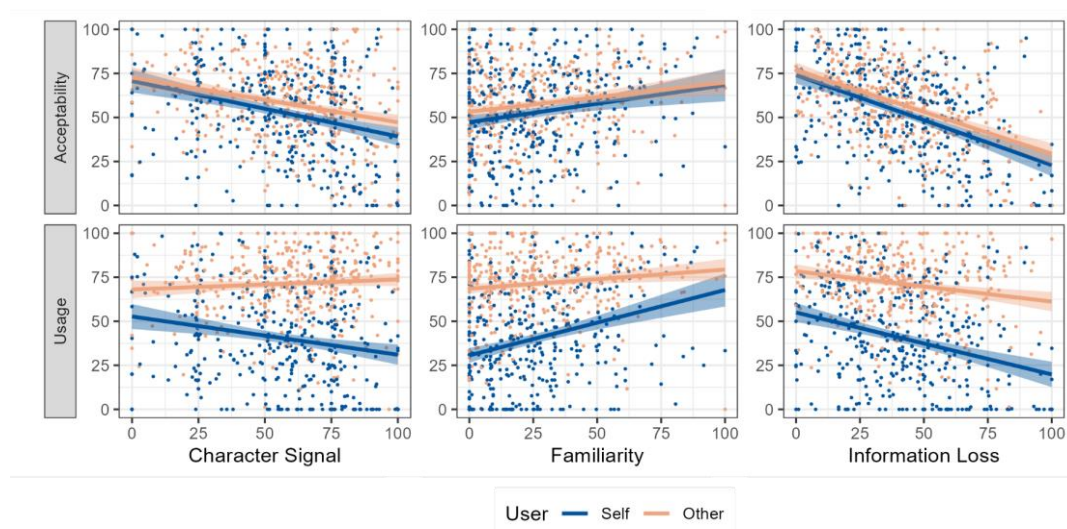

Figure A. *For perceived acceptability, Study 2 revealed negative associations to character signal, and information loss a positive association with familiarity. Regarding usage, for own*

*AICT use, the study revealed negative relationships between usage expectations and character signal, and usage expectations and information loss, and a positive association between fuselage expectations and familiarity.*

These results are expanded upon here:

- ***Acceptability and familiarity.*** We employed a linear mixed model predicting 'usage' from user (self, other), transparency (open, secret), and familiarity [0 to 100] and all possible interactions between the three predictors. Participant and medium (text voice, video) were included as random effects. With the intercept at the mean: the effect of transparency was statistically non-significant and positive ($B = 2.88$, 95% CI [-1.19, 6.95], $t(2284) = 1.39$, $p = .165$); the effect of user was statistically significant and positive ($B = 5.79$, 95% *CI* [1.72, 9.86], $t(2284) = 2.79$, $p = .005$); the effect of familiarity was statistically significant and positive ($B = 0.15$, 95% *CI* [0.11, 0.20], $t(2284) = 6.94$, $p < .001$); the interaction effect of user by transparency was statistically non-significant and positive ($B = 0.80$, 95% *CI* [-7.34, 8.94], $t(2284) = 0.19$, $p = .847$); the interaction effect of familiarity by transparency was statistically non-significant and positive ($B = 8.96e-03$, 95% *CI* [-0.07, 0.09], $t(2284) = 0.21$, $p = 0.832$); the interaction effect of familiarity by user is statistically non-significant and negative ($B = -0.04$, 95% *CI* [-0.12, 0.04], $t(2284) = -0.92$, $p = 0.356$); the interaction effect of familiarity by (transparency * user) was statistically non-significant and negative ($B = -0.09$, 95% *CI* [-0.26, 0.07], $t(2284) = -1.10$, $p = 0.271$).

- ***Acceptability and information loss.*** We employed a linear mixed model predicting 'usage' from user (self, other), transparency (open, secret), and information loss [0 to 100] and all possible interactions between the three predictors. Participant and medium (text voice, video) were included as random effects. With the intercept at the mean: the effect of transparency was statistically non-significant and negative ($B = -$ 1.59, 95% *CI* [-6.16, 2.98], *t*(2284) = -0.68, *p* = .496); the effect of user was statistically non-significant and positive (*B* = 4.52, 95% *CI* [-0.05, 9.09], *t*(2284) = 1.94, *p* = .053); the effect of information loss was statistically significant and negative (*B* = -0.41, 95% *CI* [-0.46, -0.37], *t*(2284) = -19.03, *p* < .001); the interaction effect of user by transparency was statistically non-significant and negative (B = -3.64, 95% CI [-12.78, 5.50], t(2284) = -0.78, p = 0.435); the interaction effect of information loss by transparency is statistically significant and positive (B = 0.11, 95% CI [0.03, 0.19], t(2284) = 2.57, p = 0.010); the interaction effect of information loss by user is statistically non-significant and negative (B = -2.62e-03, 95% CI [-0.08, 0.08], t(2284) = -0.06, p = 0.950); the interaction effect of information loss by (transparency * user) is statistically non-significant and positive (B = 0.07, 95% CI [-0.10, 0.23], t(2284) = 0.79, p = 0.429; Std. B = 0.01, 95% CI [-0.02, 0.05]). To explore the significant interaction between transparency and information loss we ran a linear mixed model predicting acceptability from information loss with participant and medium as random effects. We then examined secret AICTs separately to open AICTs; the effect of information loss was stronger for secret than open AICTs.

- For secret AICTs, the effect of information loss was statistically significant and negative (B = -0.47, 95% CI [-0.53, -0.41], t(1156) = -15.73, p < .001).

- For open AICTs, the effect of information loss was statistically significant and negative (B = -0.37, 95% CI [-0.43, -0.30], t(1129) = -11.46, p < .001).

- ***Acceptability and character signal.*** We employed a linear mixed model predicting 'usage' from user (self, other), transparency (open, secret), and character signal [0 to 100] and all possible interactions between the three predictors. Participant and medium (text voice, video) were included as random effects. With the intercept at the mean: the effect of transparency was statistically non-significant and positive (*B* =

3.13, 95% *CI* [-3.25, 9.51], *t*(2284) = 0.96, *p* = .336); the effect of user was statistically non-significant and positive (*B* = 4.46, 95% *CI* [-1.92, 10.83], *t*(2284) = 1.37, *p* = .171); the effect of character signal was statistically significant and negative (*B* = -0.36, 95% *CI* [-0.41, -0.31], *t*(2284) = -14.36, *p* < .001); the interaction effect of user by transparency was statistically non-significant and negative (*B* = -9.51, 95% *CI* [-22.26, 3.25], *t*(2284) = -1.46, *p* = .144); the interaction effect of character by transparency was statistically non-significant and negative (*B* = -9.96e-04, 95% *CI* [-0.09, 0.09], *t*(2284) = -0.02, *p* = .983); the interaction effect of character by agent was statistically non-significant and positive (*B* = 0.02, 95% *CI* [-0.07, 0.12], *t*(2284) = 0.45, *p* = 0.650); the interaction effect of character by (transparency * agent) is statistically non-significant and positive (*B* = 0.13, 95% *CI* [-0.06, 0.32], *t*(2284) = 1.37, *p* = .171).

**Impact on Usage expectations.** Greater usage expectations were associated with a) greater familiarity and b) lower information loss. Whereas lower usage expectations were associated with stronger character signals. These associations were stronger for participants in the 'self' condition, that is, when people were considering their own potential AICT use (see Figure A). These results were confirmed using three linear mixed effects models, one for each exploratory factor, predicting 'usage' from user (self, other), transparency (open, secret), and the exploratory factor [<familiarity/information loss/character signal>; 0 to 100]. Participant and medium were included as random effects. These results are expended upon here:

- *Usage expectations and familiarity.* We employed a linear mixed model predicting 'usage' from user (self, other), transparency (open, secret), and familiarity [0 to 100] and all possible interactions between the three predictors. Participant and medium (text voice, video) were included as random effects. With the model's intercept at the mean: the effect of transparency was statistically non-significant and positive (*B* =

0.89, 95% *CI* [-3.02, 4.80], *t*(2284) = 0.45, *p* = .656); the effect of user was statistically significant and positive (*B* = 34.71, 95% *CI* [30.80, 38.62], *t*(2284) = 17.40, *p* < .001); the effect of familiarity was statistically significant and positive (*B* = 0.19, 95% *CI* [0.15, 0.23], *t*(2284) = 8.73, *p* < .001); the interaction effect of user by transparency was statistically non-significant and positive (*B* = 3.25, 95% *CI* [-4.57, 11.07], *t*(2284) = 0.82, *p* = .415); the interaction effect of familiarity by transparency was statistically non-significant and negative (*B* = -0.02, 95% *CI* [-0.10, 0.06], *t*(2284) = -0.54, *p* = .590); the interaction effect of familiarity by user was statistically significant and negative (*B* = -0.14, 95% *CI* [-0.22, -0.06], *t*(2284) = -3.49, *p* < .001); the interaction effect of familiarity on (transparency * user) was statistically non-significant and negative (*B* = -0.15, 95% *CI* [-0.31, 0.01], *t*(2284) = -1.82, *p* = .070). To explore the interaction between familiarity and agent, we examined the effect of familiarity on usage for self and other separately. The effect of familiarity on usage was stronger for self than for other:

- For self, the effect of familiarity is statistically significant and positive (*B* = 0.22, 95% *CI* [0.15, 0.29], *t*(1126) = 6.26, *p* < .001).
- For other, The effect of familiarity is statistically significant and positive (*B* = 0.13, 95% *CI* [0.09, 0.18], *t*(1159)= 5.49, *p* < .001).

- ***Usage expectations and information loss.*** We employed a linear mixed model predicting 'usage' from user (self, other), transparency (open, secret), and information loss [0 to 100] and all possible interactions between the three predictors. Participant and medium (text voice, video) were included as random effects. With the model's intercept at the mean: the effect of transparency was statistically non-significant and negative (*B* = -1.97, 95% *CI* [-6.68, 2.75], *t*(2284) = -0.82, *p* = .414); the effect of user was statistically significant and positive (*B* = 19.67, 95% *CI* [14.96, 24.39], *t*(2284) =

8.18, $p < .001$); the effect of information loss was statistically significant and negative ($B = -0.23$, 95% $CI$ [-0.27, -0.18], $t(2284) = -10.28$, $p < .001$); the interaction effect of user by transparency is statistically non-significant and positive ($B = 1.90$, 95% $CI$ [-7.53, 11.33], $t(2284) = 0.39$, $p = .693$); the interaction effect of information loss by transparency was statistically non-significant and positive ($B = 0.05$, 95% $CI$ [-0.03, 0.13], $t(2284) = 1.19$, $p = .235$); the interaction effect of information loss by user was statistically significant and positive ($B = 0.27$, 95% $CI$ [0.19, 0.35], $t(2284) = 6.40$, $p < .001$); the interaction effect of information loss by (transparency * user) was statistically non-significant and negative ($B = -0.05$, 95% $CI$ [-0.22, 0.12], $t(2284) = -0.58$, $p = .559$).

- ***Usage expectations and character signal.*** We employed a linear mixed model predicting 'usage' from user (self, other), transparency (open, secret), and character signal [0 to 100] and all possible interactions between the three predictors. Participant and medium (text voice, video) were included as random effects. With the model's intercept at the mean: the effect of transparency was statistically non-significant and positive ($B = 5.48$, 95% CI [-0.86, 11.83], $t(2284) = 1.69$, $p = .090$); the effect of user was statistically significant and positive ($B = 9.84$, 95% CI [3.49, 16.18], $t(2284) = 3.04$, $p = .002$); the effect of character was statistically significant and negative ($B = -0.20$, 95% $CI$ [-0.25, -0.16], $t(2284) = -8.26$, $p < .001$); the interaction effect of user by transparency was statistically non-significant and negative ($B = -3.56$, 95% $CI$ [-16.25, 9.13], $t(2284) = -0.55$, $p = .582$); the interaction effect of character by transparency was statistically non-significant and negative ($B = -0.09$, 95% $CI$ [-0.18, 5.53e-03], $t(2284) = -1.84$, $p = .065$); the interaction effect of character by user was statistically significant and positive ($B = 0.37$, 95% $CI$ [0.28, 0.47], $t(2284) = 7.90$, $p < .001$); the interaction effect of character on (transparency * user) was statistically non-significant and positive ($B = 0.05$, 95% *CI* [-0.14, 0.24], $t(2284) = 0.53$, $p = 0.594$).